\documentclass[letterpaper,twocolumn,english,prb,aps,superbib,tightenlines,floatfix]{revtex4-1}
\usepackage[hypertex]{hyperref}
\usepackage{natbib}
\usepackage{graphicx,epsfig}
\usepackage{amsmath}

\begin{document}

\newcommand{\half}{\frac{1}{2}}
\newcommand{\boldsigma}{${\boldmath$\sigma$}${}}
\title{Complete All-Optical Quantum Control of Electron Spins in InAs/GaAs Quantum Dot Molecule}
\date{\today}

\author{Guy Z. Cohen}
\affiliation{Department of Physics, University of California, San
Diego, La Jolla, California 92093-0319}

\begin{abstract}
The spin states of electrons and holes confined in InAs quantum
dot molecules have recently come to fore as a promising system for
the storage or manipulation of quantum information. We describe
here a feasible scheme for complete quantum optical control of two
electron spin qubits in two vertically-stacked singly-charged InAs
quantum dots coupled by coherent electron tunneling. With an
applied magnetic field transverse to the growth direction, we
construct a universal set of gates that corresponds to the
possible Raman transitions between the spin states. We detail the
procedure to decompose a given two-qubit unitary operation, so as
to realize it with a successive application of up to 8 of these
gates. We give the pulse shapes for the laser pulses used to
implement this universal set of gates and demonstrate the
realization of the two-qubit quantum Fourier transform with
fidelity of 0.881 and duration of 414 ps. Our proposal therefore
offers an accessible path to universal computation in quantum dot
molecules and points to the advantages of using pulse shaping in
coherent manipulation of optically active quantum dots to mitigate
the negative effects of unintended dynamics and spontaneous
emission.
\end{abstract}

\maketitle

\section{Introduction} \label{introduction}

Spins in semiconductor quantum dots\cite{Yamamoto09,Liu10} (QD)
are promising candidates for satisfying the DiVincenzo criteria
for quantum computation,\cite{DiVincenzo00} such as state
initialization,\cite{Atature06,XuEmary07,Xu07,Xu08} coherent spin
manipulation,\cite{Emary07-2,Schaibley13-2} and qubit-specific
measurements.\cite{Atature07,Vamivakas10} The advantages of QD
spin qubits over other qubit realizations are numerous. Quantum
dots are compatible with existing semiconductor technology and can
be grown in millions in a regular 2D array on a single
millimeter-sized chip.\cite{Schneider08} QD spin qubits can be
integrated on a chip with photonic crystal
 cavities,\cite{Winger08,Gallo08,Carter13} have short optical
 recombination and photon emission times,\cite{Pelton02,Moreau02}
can be manipulated by fast
single-qubit\cite{Emary07-2,Greilich09,Kim10-2} and
two-qubit\cite{Chen12,Solenov13} quantum gates, and can be
entangled with adjacent qubits by tunneling
 interaction\cite{Kim10} or with remote ones via entanglement
swapping with photons.\cite{Gao12,Schaibley13,Webster14} Recently,
great progress was made in developing methods to increase the QD
spin decoherence time either through spin echo\cite{Press10} or
suppression of nuclear-spin fluctuations.\cite{Sun12}

The study of the interaction mechanisms between QDs can lead to
finding new ways to successfully manipulate their quantum state,
an important goal of quantum information technologies, like
quantum computing and quantum cryptography. Such coupling can be
obtained in optically active self-assembled
QDs,\cite{Doty08,Scheibner09} with techniques existing for
ultrafast laser initialization,\cite{Kim08}
measurement\cite{Kim08,Vamivakas10} and coherent
manipulation\cite{Emary07,Robledo08,Economou08,Kim10,Chen12} of
the spin qubits in these QDs. Many coupling mechanisms of
optically active QDs were studied including electron
tunneling,\cite{Robledo08,Kim10} hole
tunneling,\cite{Climente08,Chen12} exciton-mediated
interaction\cite{Emary07} and electron-hole exchange
interaction.\cite{Economou08} In particular, it was found that
embedding two coupled QDs, which form a QD molecule (QDM), in a
Schottky diode structure enables the tuning of the relative energy
levels of the two dots.\cite{Krenner05,Ortner05} Like real
molecules, QDMs can display bonding and anti-bonding states, that
are symmetric and anti-symmetric superpositions of localized
states.\cite{Doty08} However, unlike natural diatomic molecules,
which always have a bonding ground state, QDMs can be tailored to
have molecular ground states with anti-bonding
character.\cite{Doty09}

With the QDM embedded in a Schottky diode, the tuning of the
voltage bias determines the stable ground state charge
configuration.\cite{Scheibner09} Especially important is the case
of a doubly charged QDM,\cite{Emary07,Economou08,Kim10,Chen12}
where each of the QDs is charged with a single electron or hole,
since then the QDM can be used as a double spin qubit system. An
ambitious goal in quantum computation is complete quantum control
of such a multiple-qubit system. This control can be achieved by a
universal set of quantum gates, i.e. a set of gates with which an
arbitrary unitary operation can be constructed. Many universal
sets of quantum gates were proposed. Ref.~\onlinecite{Reck94}
showed the set of all two-level unitary gates is universal. Then,
Ref.~\onlinecite{Barenco95} showed this set of gates can be
implemented by combinations of single-qubit gates and the
two-qubit controlled-NOT (CNOT) gates. Other universal sets
consisting of only two-qubit gates were also
found.\cite{DiVincenzo95,Barenco95-2,Lloyd95,Deutsch95}

In this work, we present a scheme for complete quantum control of
a two-qubit system, a QDM composed of two vertically-stacked
singly-charged InAs QDs separated by a GaAs/AlGaAs tunnel barrier
and embedded in a Schottky diode. The two electrons interact by
kinetic exchange, which is based on coherent electron tunneling
and evidenced by a singlet-triplet splitting of the QDM energy
levels. Quantum optical control is realized through real or
virtual excitation of the bottom QD electron by laser pulses. The
universal set of gates consists of the 5 two-level unitary
operations that correspond to the possible Raman transitions in
the Voigt geometry. We find that every two-qubit unitary operation
can be realized through at most 8 such Raman transitions, and that
pulse shaping the laser pulses can substantially increase the
fidelity of the operation. As a case in point, we show a
reasonably high fidelity realization of the two-qubit quantum
Fourier transform (QFT) in this system taking light hole mixing,
decay and decoherence into account. The results indicate the
promise of QDMs as a platform for quantum computation and the
potential of pulse shaping for increasing the fidelity of quantum
state manipulation in optically active QDs.

The paper is organized as follows. In Sec.~\ref{section1} we
present the system model and Hamiltonian. We write the Hamiltonian
relative to an orthogonal basis, find the eigenstates and
eigenenergies, and plot the allowed transitions in the Faraday and
Voigt geometries. In Sec.~\ref{section2} we describe a universal
set of gates for the QDM. We give an algorithm for decomposing a
given arbitrary two-qubit unitary operation to a product of
operations realized with these gates and give criteria for a
subset of these gates to still be universal. In
Sec.~\ref{section3} we present the Hamiltonian of the interaction
with the electromagnetic field and discuss the pulse shapes used
to implement Raman transitions and the effects of light hole
mixing. In Sec.~\ref{section4} we consider the effect of decay and
decoherence. We write the Lindblad equation for the system and
derive the expression for the fidelity of an operation relative to
an ideal one. In Sec.~\ref{section5} we show, for realistic
parameters, how to apply the scheme to realize the two-qubit QFT
with reasonably high fidelity. Finally, in Sec.~\ref{conclusions}
we discuss the key results and consider directions for future
research.

\section{System Model and Hamiltonian} \label{section1}

\begin{figure}[t!]
\begin{center}
\includegraphics[width=8.5cm]{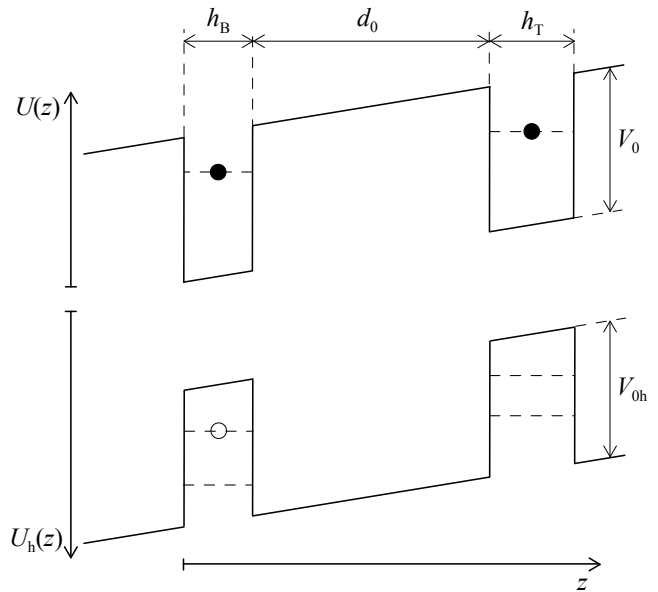}\\
\end{center}
\caption{\label{fig01} Schematic diagram of the band structure of
two vertically stacked self-assembled quantum dots embedded in a
Schottky diode. The growth direction is the $z$ direction. The
height of the bottom/top dot is $h_B$/$h_T$, while the height of
the interdot tunnel barrier is $d_0$. Optical control of the
two-electron spin state is achieved through real or virtual
transitions to the optically excited states using laser fields
tuned to create an exciton at the bottom dot only. The electric
field in the $z$ direction, $F$, is exaggerated for clarity.}
\end{figure}
The system under consideration consists of two InAs vertically
stacked self-assembled quantum dots separated by a GaAs/AlGaAs
tunnel barrier and embedded in a Schottky diode. The voltage of
the diode is adjusted to the charge stability region where each
dot occupied with a single electron at the ground state. The
coherent tunnel coupling of the electrons in the two dots is
manifested in inter-electron kinetic exchange interaction, which
gives rise to singlet and triplet electron states delocalized over
both dots. The quantum states of the electrons are optically
manipulated via laser fields that generate real or virtual
electron-hole pairs (excitons) at the bottom dot. The magnetic
field is initially taken as zero.

With the growth direction taken to be in the $z$ direction, the
potential experienced by a single electron in the conduction band
is given by
\begin{equation}
U(\mathbf{r})=\half m_\bot\omega_\bot^2(x^2+y^2)+U(z),\label{eq01}
\end{equation}
where $m_\bot$ is the effective electron mass in the $x$ and $y$
directions, approximated as independent of $z$, $\omega_{\bot}$ is
the effective frequency of the parabolic confinement in the $x$
and $y$ directions, and $U(z)$ is plotted in Fig.~\ref{fig01}. The
Hamiltonian for the $i$th electron, in turn, reads
\begin{equation}
h_i=\frac{p_{i,x}^2+p_{i,y}^2}{2m_{\bot}}+\frac{p_{i,z}^2}{2m(z_i)}+U(\mathbf{r}_i),\label{eq02}
\end{equation}
where $\mathbf{r}_i$ and $\textbf{p}_{i}$ ($i=1,2$) are the
position and momentum of the $i$th electron, and where the
effective mass in the $z$ direction is $m(z)$. The two-electron
Hamiltonian is constructed by writing Eq.~(\ref{eq02}) for each
electron and adding the Coulomb interaction term. It is
\begin{equation}
\mathcal{H}=h_1+h_2+\frac{e_0^2}{|\mathbf{r}_1-\mathbf{r}_2|},\label{eq03}
\end{equation}
where $e_0^2\equiv
e^2/(4\pi\epsilon_0)/\{[\epsilon(z_1)+\epsilon(z_2)]/2\}$, $e$
being the electron charge, $\epsilon_0$ the vacuum permittivity
and $\epsilon(z)$ the relative dielectric constant.

We consider the two dots to be separated such as to justify a
variational treatment in terms of atomic-like single-particle
states localized at the individual dots. These single particle
states are $|B\rangle|\uparrow\rangle$,
$|B\rangle|\downarrow\rangle$, $|T\rangle|\uparrow\rangle$ and
$|T\rangle|\downarrow\rangle$, where $|B\rangle$/$|T\rangle$ is a
state corresponding to the solution of the Schr\"{o}dinger
equation for the single-particle Hamiltonian in Eq.~(\ref{eq02})
with $U(z)$ containing only the bottom/top dot potential well.
Listing the two-electron combinations of these states gives the
product state basis: $|\uparrow,\uparrow\rangle$,
$|\uparrow,\downarrow\rangle$, $|\downarrow,\uparrow\rangle$,
$|\downarrow,\downarrow\rangle$, $|\downarrow\uparrow,0\rangle$
and $|0,\downarrow\uparrow\rangle$, where
$|\sigma_B,\sigma_T\rangle$ is the state with the bottom/top dot
occupied by an electron with spin projection $\sigma_B/\sigma_T$
in the $z$ direction, 0 denotes an unoccupied dot, and
$\downarrow\uparrow$ denotes a doubly-occupied dot.

With foresight, we choose a new basis composed of linear
combinations of the product state basis states. The new basis
reads
\begin{eqnarray}
|+\rangle&=&2^{-1/2}(|\downarrow\uparrow,0\rangle+|0,\downarrow\uparrow\rangle),\label{eq04}\\%bonding
|S\rangle&=&2^{-1/2}(|\uparrow,\downarrow\rangle-|\downarrow,\uparrow\rangle),\label{eq05}\\%bonding
|T_0\rangle&=&2^{-1/2}(|\uparrow,\downarrow\rangle+|\downarrow,\uparrow\rangle),\label{eq06}\\%anti-bonding
|T_+\rangle&=&|\uparrow,\uparrow\rangle,\label{eq07}\\%anti-bonding
|T_-\rangle&=&|\downarrow,\downarrow\rangle,\label{eq08}\\%anti-bonding
|-\rangle&=&2^{-1/2}(|\downarrow\uparrow,0\rangle-|0,\downarrow\uparrow\rangle),\label{eq09}%bonding
\end{eqnarray}
where $|\pm\rangle$ are the symmetric/antisymmetric doubly
occupied states, $|S\rangle$ and $|T_0\rangle$ are the singlet and
triplet Heitler-London states, and $|T_\pm\rangle$ are the two
other triplet states. The spatial parts of the triplet states are
antisymmetric, while those parts in the rest of the states are
symmetric. This corresponds to the former/latter being
anti-bonding/bonding states.

The Hamiltonian in Eq.~(\ref{eq03}) in the basis of
Eqs.~(\ref{eq04}-\ref{eq09}) is
\begin{widetext}
\begin{equation}
\mathcal{H}=(\epsilon_B+\epsilon_T)\widehat{I}+\left ( \begin{array}{cccccc} \frac{U_B+U_T}{2}+J & t_B+t_T & 0 & 0 & 0 & \epsilon_B-\epsilon_T+\frac{U_B-U_T}{2} \\
t_B+t_T & U_{BT}+J &
0 & 0 & 0 & t_B-t_T \\ 0 & 0 & U_{BT}-J & 0 & 0 & 0 \\ 0 & 0 & 0 & U_{BT}-J & 0 & 0 \\
0 & 0 & 0 & 0 & U_{BT}-J & 0 \\
\epsilon_B-\epsilon_T+\frac{U_B-U_T}{2} & t_B-t_T & 0 & 0 & 0 &
\frac{U_{B}+U_T}{2}-J
\end{array} \right ),\label{eq10}
\end{equation}
\end{widetext}
where $\widehat{I}$ is the identity matrix, and the following
definitions are employed:
\begin{eqnarray}
\epsilon_B&=&\langle B|h|B\rangle,\label{eq11}\\
\epsilon_T&=&\langle T|h|T\rangle,\label{eq12}\\
t&=&-\langle B|h|T\rangle,\label{eq13}\\
t_B&=&t-u_{BBBT}-\langle B|T\rangle\epsilon_B,\label{eq14}\\
t_T&=&t-u_{TTTB}-\langle B|T\rangle\epsilon_T,\label{eq15}\\
J&=&u_{BTBT}-2\langle B|T\rangle t.\label{eq16}
\end{eqnarray}
The single-particle Hamiltonian $h$ in
Eqs.~(\ref{eq11}-\ref{eq13}) is given in Eq.~(\ref{eq02}). The
matrix elements of this Hamiltonian in
Eqs.~(\ref{eq11}-\ref{eq12}) are the single-particle energies for
the states $|B\rangle$ and $|T\rangle$, apart from small
contributions from the other quantum well in $U(z)$. The matrix
element $t$ in Eq.~(\ref{eq13}) is termed the tunneling matrix
element. The Coulomb integral $u_{ijkl}$ ($i,j,k,l=B$ or $T$) is
given by
\begin{equation}
u_{ijkl}=\int
d\mathbf{r}_1d\mathbf{r}_2\frac{e_0^2}{|\mathbf{r}_1-\mathbf{r}_2|}\phi_i^*(\mathbf{r}_1)\phi_j^*(\mathbf{r}_2)
\phi_k(\mathbf{r}_2)\phi_l(\mathbf{r}_1),\label{eq17}
\end{equation}
with $\phi_i(\mathbf{r})$ being the atomic-like single-particle
wave function of the state $|i\rangle$. For the direct Coulomb
integrals we use the abbreviations $U_{ij}=u_{ijji}$ and
$U_i=u_{iiii}$.

\begin{figure}[t!]
\begin{center}
\includegraphics[width=8cm]{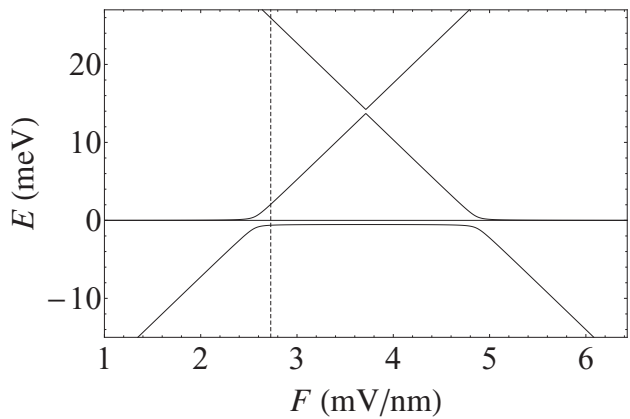}\\
\end{center}
\caption{\label{fig02} Calculated spin state energy levels for the
quantum dot molecule system vs. $F$, the applied electric field in
$z$ direction, with the experimental parameters of of $h_B=2.6$
nm, $h_T=3.2$ nm, $d_0=9$ nm, $\Delta_{\mathrm{S-T}}=$125 $\mu$eV
and zero magnetic field. The plotted energy levels at the working
point ($F=2.73$ mV/nm), denoted by a dashed line, correspond, from
bottom to top, to the modified singlet state $|\tilde{S}\rangle$;
the degenerate triplet states $|T_-\rangle$, $|T_0\rangle$ and
$|T_+\rangle$ and two linear combinations of the two doubly
occupied states $|\downarrow\uparrow,0\rangle$ and
$|0,\downarrow\uparrow\rangle$ with small (5.4\% and 0.05\% at the
working point) singlet state admixtures. The lowest/highest energy
levels were shifted down/up by 0.5 meV for the energy gaps to
become visible in the plot. These energy gaps are, from top to
bottom, the anticrossing splitting of the doubly occupied states
which equals 35 $\mu$eV and the singlet-triplet splitting which is
35$ \mu$eV at the center of the splitting ($F=3.71$ mV/nm) and
125$\mu$eV at the working point.}
\end{figure}

We make the basis of states in Eqs.~(\ref{eq04}-\ref{eq09})
orthogonal by replacing $|+\rangle$ and $|S\rangle$ in
Eqs.~(\ref{eq04}-\ref{eq05}) by $(|S\rangle\pm|+\rangle)$. We then
normalize the states of the orthogonal basis and write the
Hamiltonian in Eq.~(\ref{eq10}) in the orthonormal basis. The
diagonalization of this Hamiltonian shows the three triplet states
$|T_0\rangle$, $|T_\pm\rangle$ are degenerate eigenstates with an
energy independent of the electric field $F$. The singlet state
$|S\rangle$ is shifted in energy by the kinetic exchange
interaction below the triplet states, and this interaction also
results in an admixture of the doubly occupied states
$|\pm\rangle$ in the eigenstate dominated by $|S\rangle$. For the
specific experimental values of $h_B=2.6$ nm, $h_T=3.2$ nm,
$d_0=9$ nm (see Fig.~\ref{fig01}) and $\Delta_{\mathrm{S-T}}=125$
$\mu$eV, taken from Ref.~\onlinecite{Kim10}, we plot the
eigenenergies of $\mathcal{H}$ as a function of the external
electric field $F$ in Fig.~\ref{fig02}. The working point is taken
as the one for which the singlet-triplet splitting is
$\Delta_{\mathrm{S-T}}=125$ $\mu$eV, and at this point the singlet
eigenstate is numerically obtained as
$|\tilde{S}\rangle=0.973|S\rangle-0.173|+\rangle+0.149|-\rangle$.
We term this state together with the triplet states, the spin
states.

We now consider the optically excited states in our system, namely
the $X^{2-}$ states which consist of a negative trion at the
bottom dot and an unpaired electron at the top dot. These states
are $|\downarrow\uparrow\Uparrow,\uparrow\rangle$,
$|\downarrow\uparrow\Uparrow,\downarrow\rangle$,
$|\downarrow\uparrow\Downarrow,\uparrow\rangle$, and
$|\downarrow\uparrow\Downarrow,\downarrow\rangle$, with
$|\downarrow\uparrow\Uparrow\rangle=2^{-1/2}(|\uparrow\downarrow\rangle-|\downarrow\uparrow\rangle)|\Uparrow\rangle$
and
$|\downarrow\uparrow\Downarrow\rangle=2^{-1/2}(|\uparrow\downarrow\rangle-|\downarrow\uparrow\rangle)|\Downarrow\rangle$
representing trions, and with
$|\Uparrow\rangle=|\frac{3}{2},\frac{3}{2}\rangle$ and
$|\Downarrow\rangle=|\frac{3}{2},-\frac{3}{2}\rangle$ being the
heavy hole states with $3/2$ and $-3/2$ spin projections along
$z$. The spin states can be excited to the optically excited
states, which we term the trion states, through the creation of an
exciton at the bottom dot by the laser field. The 4 trion states
have the same single-particle and Coulomb interaction terms in
their energies. Moreover, we calculated for a wide range of
relevant experimental values that the electron-hole spin exchange
interaction is negligible, since the unpaired electron and hole
reside in different dots. We also found that hole tunneling is too
weak to have a notable effect on the energies due to the highly
localized hole wave functions. We therefore conclude the trion
states are degenerate.
\begin{figure}[t!]
\begin{center}
\includegraphics[width=8.5cm]{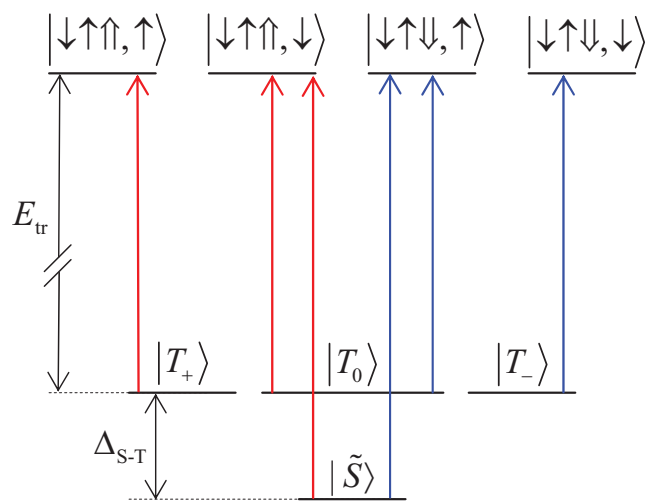}\\
\end{center}
\caption{\label{fig03} (color online) Energy level diagram of the
quantum dot molecule in the Faraday geometry with zero magnetic
field. The states include the modified singlet state
$|\tilde{S}\rangle$; the degenerate triplet states $|T_0\rangle$,
$|T_+\rangle$ and $|T_-\rangle$; and the optically excited
degenerate trion states
$|\downarrow\uparrow\Uparrow,\uparrow\rangle$,
$|\downarrow\uparrow\Uparrow,\downarrow\rangle$,
$|\downarrow\uparrow\Downarrow,\uparrow\rangle$ and
$|\downarrow\uparrow\Downarrow,\downarrow\rangle$ with energy
$E_{tr}$. The kinetic exchange interaction shifts the modified
singlet energy level below that of the triplet states with the
splitting magnitude being $\Delta_{S-T}$. The selection rules are
shown in the diagram with the red/blue arrows denoting
$\sigma_\pm$ polarization.}
\end{figure}
\begin{figure}[t!]
\begin{center}
\includegraphics[width=8.5cm]{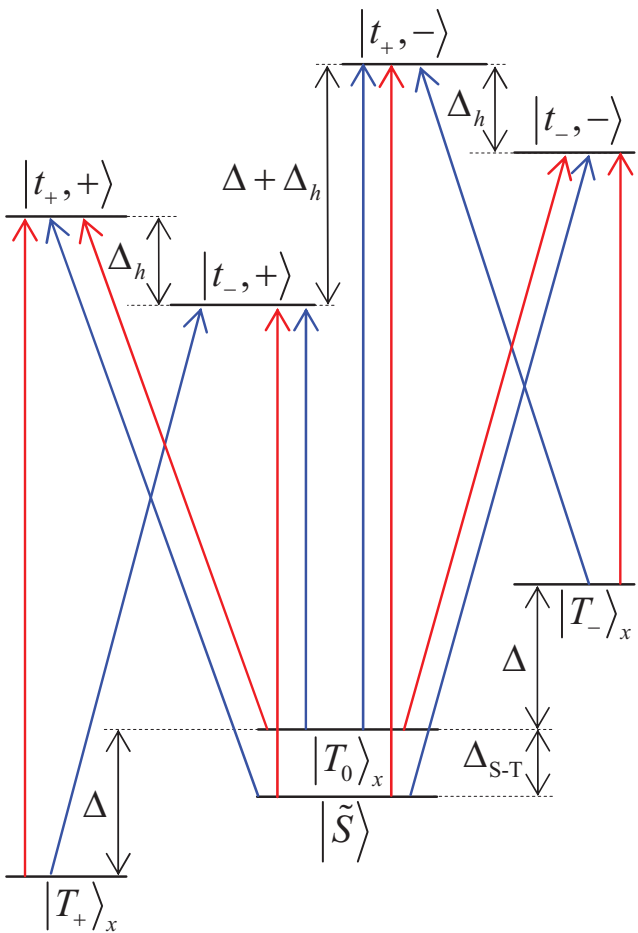}\\
\end{center}
\caption{\label{fig04} (color online) Energy level diagram of the
quantum dot molecule in the Voigt geometry (growth direction is
$z$, and magnetic field is in $+x$ direction). The energy of the
modified singlet state $|\tilde{S}\rangle$ is unaffected by the
magnetic field and remains split from $|T_0\rangle_x$ by
$\Delta_{\mathrm{S-T}}$, while the triplet states are Zeeman-split
in energy to $|T_+\rangle_x$, $|T_0\rangle_x$ and $|T_-\rangle_x$,
with the splitting being $\Delta=|g_e|\mu_BB$, $g_e$ the electron
$g$-factor, $\mu_B$ the Bohr magnetron and $B$ the magnitude of
the magnetic field. The degeneracy of the optically excited states
is also removed by the magnetic field and the states are
Zeeman-split as shown in the figure, where $\Delta_h=|g_h|\mu_B B$
and $g_h$ is the hole $g$-factor. The selection rules are plotted
in the diagram with red arrows corresponding to $V$ ($\pi_y$)
polarization, and blue arrow arrows to $H$ ($\pi_x$)
polarization.}
\end{figure}

With the energies of the spin states and the trion states
determined, we obtain the selection rules and plot the energy
level diagram for the system in the Faraday geometry and zero
magnetic field in Fig.~\ref{fig03}. The allowed transitions are
with circularly polarized light with polarizations $\sigma_\pm$.
We notice the $|T_\pm\rangle$ states are each isolated from the
rest of the spin states in the sense that no Raman transition can
be induced between each of these states and another spin state.
This result, which persists when a magnetic field in the $z$
direction is applied, precludes complete optical control of the
two-electron spin state in this geometry and, consequently ,we
decide to consider the Voigt geometry with a magnetic field
applied in the $+x$ direction.

The application of a magnetic field in the $+x$ direction
Zeeman-splits the triplet states in energy to the states
$|T_-\rangle_x$, $|T_0\rangle_x$ and $|T_+\rangle_x$ as shown in
Fig.~\ref{fig04}, while the modified singlet state, having no spin
projection in the $x$ direction, is not shifted in energy. The new
triplet states are given by
\begin{eqnarray}
|T_-\rangle_x&=&|-,-\rangle,\label{eq18}\\
|T_0\rangle_x&=&2^{-1/2}(|+,-\rangle+|-,+\rangle),\label{eq19}\\
|T_+\rangle_x&=&|+,+\rangle,\label{eq20}
\end{eqnarray}
where
\begin{equation}
|\pm\rangle=2^{-1/2}(|\uparrow\rangle\pm|\downarrow\rangle)\label{eq21}
\end{equation}
are the eigenstates of the spin operator in $x$ direction. The
trion states are also Zeeman-split in energy, with the new states
being $|t_+,\pm\rangle$ and $|t_-,\pm\rangle$ and where the states
$|t_\pm\rangle$ are defined by
\begin{equation}
|t_\pm\rangle=2^{-1/2}(|\downarrow\uparrow\Uparrow\rangle\pm|\downarrow\uparrow\Downarrow\rangle).\label{eq22}
\end{equation}
We plot the trion states Zeeman splittings in Fig.~\ref{fig04}.
With the energy levels determined, we write the selection rules
for the Voigt geometry. We find the allowed transitions have
linear polarizations in either the $x$ or $y$ directions, and that
the set of transitions, shown in Fig.~\ref{fig04}, makes Raman
transitions possible between all pairs of spin states apart from
the pair of $|T_-\rangle_x$ and $|T_+\rangle_x$, a point that will
be important later when we tackle the problem of complete quantum
control of the two-electron spin state.

\section{Universal Computation with Two-Level Operations} \label{section2}

The spin states in the quantum dot molecule system in the Voigt
geometry form a basis for a two-qubit computational state space.
This basis is $|\pm,\pm\rangle$, where the $|\pm\rangle$ states
were defined in Eq.~(\ref{eq21}), and where the admixtures of the
doubly-occupied states were omitted for brevity. The scheme for
universal computation we propose provides a realization of a given
arbitrary unitary operation in the computational basis through the
application of a set of two-level unitary operations between the
spin states. Each of the two-level unitary operations operates on
two spin levels and is realized through two laser pulses with the
same Raman detuning $\delta$ that pump the transitions from the
two spin states to a common trion state.

Suppose we want to realize a given a unitary operation $U$ in the
computational basis as a product of two-level unitary operations.
A general proper unitary two-level operation in the space of the
states $|i\rangle$ and $|j\rangle$ ($i,j=1,\dots,4$) in the spin
state basis , $|\tilde{S}\rangle$, $|T_0\rangle_x$,
$|T_+\rangle_x$ and $|T_-\rangle_x$, is given by
\begin{equation}
R_{ij}(\theta,\widehat{n})=\exp(-i\frac{\theta}{2}\widehat{n}\cdot
\boldsymbol\sigma ),\label{eq23}
\end{equation}
which rotates the pseudo-spin vector in the subspace of the states
$|i\rangle$ and $|j\rangle$ by an angle $\theta$ about the axis
$\widehat{n}$, with
$\boldsymbol\sigma=(\sigma_x,\sigma_y,\sigma_z)$ acting in the
subspace of these two levels. Going back to the unitary operation
$U$, we first write its matrix relative to the spin state basis.
Then, Ref.~\onlinecite{Reck94} shows how to decompose an arbitrary
$n$-qubit unitary operation to a product of up to ${2^n \choose
2}$ two-level unitary operations. However, we prefer the
operations in the product to be proper unitary, as such operations
are more easily implemented experimentally,\cite{Piermarocchi02}
and we give in Appendix \ref{app-A} the procedure for decomposing
$U$ to a product of up to 6 proper unitary two-level operators in
the form of Eq.~(\ref{eq23}) and an overall phase factor, which
can be omitted.

\begin{figure}[t!]
\begin{center}
\includegraphics[width=5cm]{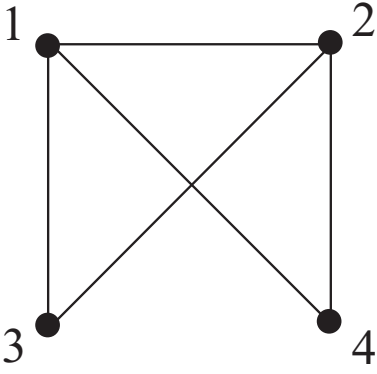}\\
\end{center}
\caption{\label{fig05} Graph representation of the spin states and
the possible two-level operations between them. The graph vertices
represent levels and its edges represent possible two-level
unitary operations. The vertices $1$, $2$, $3$ and $4$ are
associated with the states $|\tilde{S}\rangle$, $|T_0\rangle_x$,
$|T_+\rangle_x$ and $|T_-\rangle_x$, respectively.}
\end{figure}

From the energy level diagram in Fig.~\ref{fig04} it can been seen
that apart from $R_{34}$ each of the 6 two-level unitary operators
that may appear in the decomposition of $U$, namely $R_{12}$,
$R_{13}$, $R_{14}$, $R_{23}$, $R_{24}$ and $R_{34}$, may be
implemented with a single Raman transition. A graphical
illustration of the situation appears in the graph in
Fig.~\ref{fig05}, wherein each vertex represents a level and each
edge a possible two-level unitary operation. The operation
$R_{34}$ may be implemented through 3 Raman transitions as shown
by the following argument. If we represent a $\pi$ rotation
operation between the states $|i\rangle$ and $|j\rangle$ as
\begin{equation}
P_{ij}=R_{ij}(\pi,\widehat{y}),\label{eq24}
\end{equation}
then we have
\begin{eqnarray}
R_{ik}&=&P_{jk}R_{ij}P_{jk}^\dagger,\label{eq25}\\
R_{kj}&=&P_{ik}R_{ij}P_{ik}^\dagger.\label{eq26}
\end{eqnarray}

With the problem of $R_{34}$ solved, we consider the fact that it
may be advantageous to use of only some of the two-level unitary
operations, as the realizations of those using Raman transitions
may be faster or have higher fidelity relative to the other
operations. We therefore ask what universal subsets of the 5
operations can be chosen, i.e. subsets with which any arbitrary
unitary operation can be implemented. The answer to this question
lies in the graph interpretation of Eqs.~(\ref{eq25}-\ref{eq26}).
These equations show that an edge may be realized by 3 Raman
transitions if a two-edge path connects its vertices. Extending
this result by induction, we find that an edge can be realized by
$2n-1$ Raman transitions if a path of $n$ edges connects its
vertices. An important result follows. For a subset of the graph
in Fig.~\ref{fig05} to be universal, it must include all vertices
and be connected. When we define $\pi$-edges as edges representing
$\pm\pi$ rotations, the graph for the minimal universal set of
operations can be found and its form is given in App.~\ref{app-B}.

\section{Pulse Shaping} \label{section3}

The two-level unitary operations in our system are realized
through Raman transitions. For such a transition to implement
$R_{ij}$ the quantum dot molecule is illuminated by two
phase-locked laser pulses propagating in the $z$ direction. The
two pulses are linearly polarized, following the selection rules
in Fig.~\ref{fig04}, in either the vertical $(V)$ or horizontal
$(H)$ directions, have a common Raman detuning $\delta$ and pump
the two transitions from the spin states $|i\rangle$ and
$|j\rangle$ to a common trion state. The model Hamiltonian in the
basis of $|\tilde{S}\rangle$, $|T_0\rangle_x$, $|T_+\rangle_x$,
$|T_-\rangle_x$, $|t_+,+\rangle$, $|t_+,-\rangle$,
$|t_-,+\rangle$, and $|t_-,-\rangle$, in the rotating wave
approximation (RWA) is
\begin{widetext}
\begin{equation}
\mathcal{H}=\left ( \begin{array}{cccccccc}
-\Delta_{\mathrm{S-T}} & 0 & 0 & 0 & \chi^*_H & -\chi^*_V & \chi^*_V & -\chi^*_H \\
 0 & 0 & 0 & 0 & \chi^*_V & -\chi^*_H & \chi^*_H & -\chi^*_V \\
 0 & 0 & -\Delta & 0 & \sqrt{2}\chi_V^* & 0 & \sqrt{2}\chi^*_H & 0 \\
 0 & 0 & 0 & \Delta & 0 & \sqrt{2}\chi^*_H & 0 & \sqrt{2}\chi_V^* \\
 \chi_H  & \chi_V  & \sqrt{2}\chi_V & 0                 & E_{tr}+(\Delta_h-\Delta)/2 & 0 & 0 & 0 \\
 -\chi_V & -\chi_H & 0 & \sqrt{2}\chi_H & 0 & E_{tr}+(\Delta_h+\Delta)/2 & 0 & 0 \\
 \chi_V  & \chi_H & \sqrt{2}\chi_H & 0                & 0 & 0 & E_{tr}-(\Delta_h+\Delta)/2 & 0 \\
 -\chi_H & -\chi_V & 0 & \sqrt{2}\chi_V & 0 & 0 & 0 &
 E_{tr}-(\Delta_h-\Delta)/2
\end{array} \right ),\label{eq27}
\end{equation}
\end{widetext}
where $E_{tr}$ is the trion state energy shown in
Fig.~\ref{fig03}, we have set $\hbar=1$, $\Delta=|g_e|\mu_BB$ is
the electron Zeeman splitting, $\Delta_h=|g_h|\mu_BB$ is the hole
Zeeman splitting, and where $\mu_B$ is the Bohr magnetron, $B$ is
the magnetic field magnitude, and $g_e$ and $g_h$ are the electron
and hole $g$-factors, respectively. The expressions
$\chi_V(t)=\Omega_V(t)/2$ and $\chi_H(t)=\Omega_H(t)/2$ in
Eq.~(\ref{eq27}) are, respectively, halves of the time-dependent
Rabi frequency for the transitions
$|\tilde{S}\rangle\rightarrow|t_-,+\rangle$ and
$|T_0\rangle_x\rightarrow|t_-,+\rangle$.

In the interaction representation the Hamiltonian in
Eq.~$(\ref{eq27})$ is recast as
\begin{widetext}
\begin{equation}
\mathcal{H}=\left (
\begin{array}{cccccccc}
0 & 0 & 0 & 0 & \chi^*_H e^{i\Delta_{15}t} & -\chi^*_V e^{i\Delta_{16}t} & \chi^*_V e^{i\Delta_{17}t} & -\chi^*_H e^{i\Delta_{18}t} \\
 0 & 0 & 0 & 0 & \chi^*_V e^{i\Delta_{25}t} & -\chi^*_H e^{i\Delta_{26}t} & \chi^*_H e^{i\Delta_{27}t} & -\chi^*_V e^{i\Delta_{28}t} \\
 0 & 0 & 0 & 0   & \sqrt{2}\chi_V^* e^{i\Delta_{35}t} & 0 & \sqrt{2}\chi^*_H e^{i\Delta_{37}t} & 0 \\
  0 & 0 & 0 & 0 & 0 & \sqrt{2}\chi^*_H e^{i\Delta_{46}t} & 0 & \sqrt{2}\chi_V^* e^{i\Delta_{48}t} \\
\chi_H e^{-i\Delta_{15}t}   & \chi_V e^{-i\Delta_{25}t}  & \sqrt{2}\chi_V e^{-i\Delta_{35}t} & 0 & 0 & 0 & 0 & 0 \\
 -\chi_V e^{-i\Delta_{16}t} & -\chi_H e^{-i\Delta_{26}t} & 0 & \sqrt{2}\chi_H e^{-i\Delta_{46}t} & 0 & 0 & 0 & 0 \\
 \chi_V e^{-i\Delta_{17}t}  & \chi_H e^{-i\Delta_{27}t}  & \sqrt{2}\chi_H e^{-i\Delta_{37}t} & 0 & 0 & 0 & 0 & 0 \\
 -\chi_H e^{-i\Delta_{18}t} & -\chi_V e^{-i\Delta_{28}t} & 0 & \sqrt{2}\chi_V e^{-i\Delta_{48}t} & 0 & 0 & 0 & 0
\end{array} \right ),\label{eq28}
\end{equation}
\end{widetext}
where $\Delta_{ij}=E_i-E_j$, and where $E_i$ is the energy of the
state $|i\rangle$. Eqs.~(\ref{eq27}) and (\ref{eq28}) were derived
with $|\tilde{S}\rangle$ approximated as the singlet state
$|S\rangle$. When the full expression for $|\tilde{S}\rangle$ is
used, the Hamiltonian matrices are the same, apart from an extra
numerical factor in the off-diagonal matrix elements involving
$|\tilde{S}\rangle$. This factor is the coefficient of $|S\rangle$
in $|\tilde{S}\rangle$, which for the experimental values in
Ref.~\onlinecite{Kim10} is 0.973.

As seen from the energy level diagram in Fig.~\ref{fig04}, a Raman
transition from one spin state to another may suffer from
unintended dynamics due to the presence of close resonances.
Rather than increase the pulses lengths, which will make the
operation slower and magnify the effects of dephasing, we cope
with the unintended dynamics through pulse
shaping,\cite{Chen01,Piermarocchi02} whereby we vary the
parameters of the pulses to optimize the operation.

We describe the explicit form of $\chi_V(t)$ and $\chi_H(t)$ with
pulse shaping for the V-H, H-H and V-V cases, with the case names
denoting the polarizations of the Raman laser pulses involved. In
the V-H case we have
\begin{eqnarray}
\chi_V(t)&=&\chi_{0}(t)+\tilde{\chi}_{2}(t),\label{eq29}\\
\chi_H(t)&=&\chi_{1}(t-\Delta t)+\tilde{\chi}_{3}(t-\Delta
t),\label{eq30}
\end{eqnarray}
where
\begin{eqnarray}
\chi_{j}(t)&=&\chi_{j}e^{-(t/\tau_{j})^2}e^{-i\omega_{j}t+i\varphi_{j}},\label{eq31}\\
\tilde{\chi}_{j}(t)&=&\chi_{j}(t)e^{i\varphi_{j}'\ln\cosh(t/\tau_{j})},\label{eq32}
\end{eqnarray}
and where $\chi_{j}(t)$ ($j=0,1$) is a main pulse with amplitude
$\chi_{j}$, pulse width $\tau_{j}$, central frequency $\omega_{j}$
and constant phase $\varphi_{j}$. The shift in time between the
two main pulses is given by $\Delta t$. The expression
$\tilde{\chi}_{j}(t)$ ($j=2,3$) in Eq.~(\ref{eq32}) is a helping
pulse designed to alleviate the effects of unintended dynamics.
This pulse has an additional chirping term,\cite{Liu05} which
sweeps the instantaneous frequency in time through a range of
frequencies centered about $\omega_{j}$.

The pulse shapes for the other two cases involve only one
polarization type. In the H-H case we have
\begin{eqnarray}
\chi_V(t)&=&0,\label{eq33}\\
\chi_H(t)&=&\chi_{0}(t)+\tilde{\chi}_{2}(t)+\nonumber\\
&&\chi_{1}(t-\Delta t)+\tilde{\chi}_{3}(t-\Delta t),\label{eq34}
\end{eqnarray}
and in the V-V case we have
\begin{eqnarray}
\chi_V(t)&=&\chi_{0}(t)+\tilde{\chi}_{2}(t)+\nonumber\\
&&\chi_{1}(t-\Delta t)+\tilde{\chi}_{3}(t-\Delta
t),\label{eq35}\\
\chi_H(t)&=&0.\label{eq36}
\end{eqnarray}

The pulse parameters should be varied so as to maximize the
fidelity of the operation relative to the ideal $R_{ij}$. As this
fidelity depends on the input state, the expected value of the
fidelity over all possible input states is required. Let the
states in the basis $|\tilde{S}\rangle$, $|T_0\rangle_x$,
$|T_+\rangle_x$, $|T_-\rangle_x$, $|t_+,+\rangle$,
$|t_+,-\rangle$, $|t_-,+\rangle$, and $|t_-,-\rangle$ be denoted
by $|j\rangle$ ($j=1,\dots,8$). For unitary evolution we find in
App.~\ref{app-C} the expected fidelity is
\begin{equation}
\overline{\mathcal{F}}=\frac{1}{20}\sum\limits_{i=1}^4\sum\limits_{j=1}^4
(I_{ii}I^*_{jj}+|I_{ij}|^2),\label{eq42}
\end{equation}
where
\begin{equation}
I_{ij}=\langle i|\tilde{U}^\dagger U_{id}|j\rangle,\label{eq43}
\end{equation}
and where $\tilde{U}$ and $U_{id}$ are, respectively, the actual
and ideal evolution operators. The expected fidelity in
Eq.~(\ref{eq42}) is the yardstick by which we decide whether one
set of pulse parameters is better than another set of parameters
for the implementation of a given two-level unitary operation
$R_{ij}$.

We now incorporate the effects of light hole mixing in our
analysis. The hole states $|\Uparrow\rangle$ and
$|\Downarrow\rangle$ in Eq.~(\ref{eq22}), denoted as
$|H_z^\pm\rangle$, and heretofore taken as bare heavy hole states,
are better approximated through the Luttinger
Hamiltonian\cite{Luttinger55,Broido85} as a superposition of heavy
hole states and light hole states of the form
\begin{equation}
|H_z^\pm\rangle=\cos\theta_m|\frac{3}{2},\pm\frac{3}{2}\rangle-\sin\theta_m
e^{\mp i\phi_m}|\frac{3}{2},\mp \frac{1}{2}\rangle.\label{eq44}
\end{equation}
In Eq.~(\ref{eq44})  $\theta_m$ and $\phi_m$ are the mixing
angles, $|\frac{3}{2},\pm\frac{3}{2}\rangle$ are the heavy hole
states, $|\frac{3}{2},\pm\frac{1}{2}\rangle$ the light hole
states, and the quantization axis is taken as the growth direction
($z$ direction).

When the new hole states in Eq.~(\ref{eq44}) are substituted in
the trion states and the Hamiltonian matrix is calculated
following an analogous derivation in the single dot case in
Ref.~\onlinecite{Emary07-2}, we find results identical with the
ones obtained in this reference, namely that adjusting the
polarizations of the lasers from $\mathbf{V}$ and $\mathbf{H}$ to
$\mathbf{V}'=2^{-1/2}(\boldsymbol{\sigma}_++e^{i\mu_+}\boldsymbol\sigma_-)$
and
$\mathbf{H}'=2^{-1/2}(\boldsymbol{\sigma}_+-e^{-i\mu_-}\boldsymbol\sigma_-)$,
with
\begin{equation}
e^{i\mu_\pm}=\frac{\sqrt{3}\cos\theta_m\pm\sin\theta_me^{\pm
i\phi_m}}{\sqrt{3}\cos\theta_m\pm\sin\theta_me^{\mp
i\phi_m}},\label{eq45}
\end{equation}
the Hamiltonian has the same form as in the case with no light
hole mixing [Eq.~(\ref{eq28})] apart from replacing $\chi_V$ and
$\chi_H$ by
\begin{eqnarray}
\tilde{\chi}_{V'}&=&\chi_{V'}\frac{1+2\cos
2\theta_m}{3\cos\theta_m+\sqrt{3}e^{-i\phi_m}\sin\theta_m},\label{eq46}\\
\tilde{\chi}_{H'}&=&\chi_{H'}\frac{1+2\cos
2\theta_m}{3\cos\theta_m-\sqrt{3}e^{i\phi_m}\sin\theta_m},\label{eq47}
\end{eqnarray}
where $\chi_{V'}$ and $\chi_{H'}$ are the analogous quantities to
$\chi_V$ and $\chi_H$ for light with $V'$ and $H'$ polarizations,
respectively, and the effects of $\mathbf{V}'$ and $\mathbf{H}'$
being non-orthogonal are neglected.

The identity of the last results with the single dot case in
Ref.~\onlinecite{Emary07-2} is not surprising, as only the bottom
quantum dot is optically excited. To summarize, the effects of
light holes mixing can be circumvented by adjusting the
polarization of the laser pulses and their Rabi frequencies as
noted above. With these adjustments, the Hamiltonian in
Eq.~(\ref{eq28}) can still be used.

\section{Decay and Decoherence} \label{section4}

When decay and decoherence are taken into account the state of the
system may be described by a density matrix $\rho$ and its
non-unitary evolution by a quantum master equation in the Lindblad
form
\begin{equation}
\dot{\rho}=-i[\mathcal{H},\rho]+\sum\limits_i
\mathcal{L}_i[\rho],\label{eq48}
\end{equation}
where $H$ is given in Eq.~(\ref{eq28}), and where the sum is over
all trion state relaxation channels, each of which is described by
a Lindblad superoperator
\begin{equation}
\mathcal{L}_i[\rho]=D_i\rho D_i^\dagger-\half D_i^\dagger
D_i\rho-\half \rho D_i^\dagger D_i.\label{eq49}
\end{equation}
In taking the relaxation channels as separate and incoherent, we
neglected spontaneously generated coherence (SGC), since we
assumed the Zeeman splittings are large enough to satisfy
$|\Delta_{jk}|\gg \Gamma$ ($j,k=5,\dots,8$; $j\neq k$), where
$\Gamma$ is the total relaxation rate of a given trion
level,\cite{Economou05,Breuer07} and that the level splittings
$\Delta$, $\Delta_h$ and $\Delta_{S-T}$ are larger or of the same
order as the Rabi frequency $\Omega$.\cite{Berman11} The effects
of pure dephasing are not included in Eq.~(\ref{eq48}), as this
process, which is generated by the nuclear spins, has a rate much
lower than the total trion dephasing rate in this
system.\cite{Ulrich11,Schaibley13-2}

The 12 jump operators, $D_i$, in Eq.~(\ref{eq48}), each
corresponding to a possible relaxation channel, are
\begin{eqnarray}
D_1&=&\sqrt{\Gamma/4}|\tilde{S}\rangle\langle t_+,+|,\label{eq50}\\
D_2&=&\sqrt{\Gamma/4}|T_0\rangle_x\langle t_+,+|,\label{eq51}\\
D_3&=&\sqrt{\Gamma/2}|T_+\rangle_x\langle t_+,+|,\label{eq52}\\
D_4&=&\sqrt{\Gamma/4}|\tilde{S}\rangle\langle t_+,-|,\label{eq53}\\
D_5&=&\sqrt{\Gamma/4}|T_0\rangle_x\langle t_+,-|,\label{eq54}\\
D_{6}&=&\sqrt{\Gamma/2}|T_-\rangle_x\langle t_+,-|,\label{eq55}\\
D_7&=&\sqrt{\Gamma/4}|\tilde{S}\rangle\langle t_-,+|,\label{eq56}\\
D_8&=&\sqrt{\Gamma/4}|T_0\rangle_x\langle t_-,+|,\label{eq57}\\
D_9&=&\sqrt{\Gamma/2}|T_+\rangle_x\langle t_-,+|,\label{eq58}\\
D_{10}&=&\sqrt{\Gamma/4}|\tilde{S}\rangle\langle t_-,-|,\label{eq59}\\
D_{11}&=&\sqrt{\Gamma/4}|T_0\rangle_x\langle t_-,-|,\label{eq60}\\
D_{12}&=&\sqrt{\Gamma/2}|T_-\rangle_x\langle t_-,-|.\label{eq61}
\end{eqnarray}
The terms in the square roots in Eqs.~(\ref{eq50}-\ref{eq61}) are
the relaxation rates for the corresponding decay channels. These
rates were determined by splitting the total decay rate, the trion
state linewidth $\Gamma$, in the ratio of absolute square values
of the dipole matrix elements for the allowed transitions from the
this state to the spin states, as these values are proportional to
the spontaneous decay rates of the corresponding channels.

With spontaneous decay and decoherence included in the model, the
expression for the expected fidelity of an operation relative an
ideal operation in Eq.~(\ref{eq42}) should be generalized. The
derivation in App.~\ref{app-D} gives this generalized expected
fidelity as
\begin{equation}
\overline{\mathcal{F}}=\frac{1}{20}\sum\limits_{i=1}^4\sum\limits_{j=1}^4\biggl[\langle
j|U_{id}^\dagger\rho^{(ji)}U_{id}|i\rangle+\langle
j|U_{id}^\dagger\rho^{(ii)}U_{id}|j\rangle\biggr],\label{eq62}
\end{equation}
where $\rho^{(ij)}$ is the density matrix resulting from the
non-unitary evolution of an initial density matrix
$|i\rangle\langle j|$. In the limit of unitary evolution
($\Gamma=0$), $\rho^{(ij)}=\tilde{U}|i\rangle\langle
j|\tilde{U}^\dagger$ and Eq.~(\ref{eq62}) reduces to
Eq.~(\ref{eq42}).

\section{The Two-Qubit Quantum Fourier Transform} \label{section5}

As a case in point of a non-trivial two-qubit transformation
realized using the methods described above, we choose the
two-qubit quantum Fourier transformation. In the computational
basis $|\pm,\pm\rangle$ this transformation is
\begin{equation}
U_{QFT}=\half\left ( \begin{array}{cccc} 1 & 1 & 1 & 1 \\ 1 & i & -1 & -i \\ 1 & -1 & 1 & -1 \\
1 & -i & -1 & i \end{array} \right ).\label{eq63}
\end{equation}
Rewriting Eq.~(\ref{eq63}) relative to the physical basis
$|\tilde{S}\rangle$, $|T_0\rangle_x$, $|T_+\rangle_x$, and
$|T_-\rangle_x$ as $\widetilde{U}_{QFT}$, and applying the
algorithm in App.~\ref{app-A} to decompose the result to a product
of two-level proper unitary operations, we find
\begin{eqnarray}
\widetilde{U}_{QFT}=&&e^{i\alpha}R_{32}(\theta_{32},\widehat{n}_{32})R_{34}(\theta_{34},\widehat{n}_{34})R_{12}(\theta_{12},\widehat{n}_{12})\times\nonumber\\
&&R_{14}(\theta_{14},\widehat{n}_{14})R_{24}(\theta_{24},\widehat{n}_{24}),\label{eq64}
\end{eqnarray}
where $R_{ij}(\theta,\widehat{n})$ is defined in Eq.~(\ref{eq23}),
and where $\alpha$, $\theta_{ij}$ and $\widehat{n}_{ij}$ were
numerically determined.

We detail the optimization process of the pulse train. First, we
choose the physical parameters for the system as the ones in
Ref.~\onlinecite{Kim10}, namely $h_B=2.6$ nm, $h_T=3.2$ nm,
$d_0=9$ nm (see Fig.~\ref{fig01}) and $\Delta_{\mathrm{S-T}}=125$
$\mu$eV. The electron and hole $g$-factors are taken from
Ref.~\onlinecite{Xu07} as $g_e=-0.48$ and $g_h=0.31$, and the
trion states linewidth $\Gamma$ is obtained from
Ref.~\onlinecite{Atature06} as $\Gamma=1.2$ $\mu$eV. Next, for
each of the operations $R_{12}$, $R_{13}$, $R_{14}$, $R_{23}$ and
$R_{24}$ we choose the trion state that will be used to realize
the Raman transition. We choose this state so as to reduce the
unintended dynamics, i.e. such that the resonance frequencies of
each of the transitions from the two spin states to this trion
state is as far as possible from other allowed transitions with
the same polarization. We then choose a value for the magnetic
field $B$ that maximizes the minimum absolute frequency difference
among all pairs of allowed transitions with the same polarization.
For magnetic field values lower than 10 T, this value is found to
be $B=8.99$ T, and the minimum absolute frequency difference is
36.5 $\mu$eV.

We then turn to the optimization of the individual pulses. Since
each $R_{ij}$ in Eq.~(\ref{eq64}) (apart from $R_{34}$) can be
implemented directly as given or indirectly via
Eqs.~(\ref{eq25}-\ref{eq26}), we consider all possible indirect
implementations and optimize the 35 operations
$R_{ij}(\theta_{32},\widehat{n}_{32})$,
$R_{ij}(\theta_{34},\widehat{n}_{34})$,
$R_{ij}(\theta_{12},\widehat{n}_{12})$,
$R_{ij}(\theta_{14},\widehat{n}_{14})$,
$R_{ij}(\theta_{24},\widehat{n}_{24})$, $P_{ij}$ and
$P^\dagger_{ij}$, where $(i,j)=(1,2),(1,3),(1,4),(2,3),(2,4)$. In
each optimization we realize the operation with two main pulses,
each of which having the form of Eq.~(\ref{eq31}), and maximize
the expected fidelity in Eq.~(\ref{eq42}) through grid search in
pulse parameter space followed by gradient descent. In the
optimization process we keep in mind the need for short pulses
that will ensure fast operation. Examining the results of the
optimizations, we conclude that direct implementation is always
preferable when possible, with the only case where indirect
implementation is chosen being
\begin{eqnarray}
R_{34}(\theta_{34},\widehat{n}_{34})=P_{13}
R_{14}(\theta_{34},\widehat{n}_{34})P_{13}^\dagger,\label{eq65}
\end{eqnarray}
which cannot be directly implemented due to the selection rules.

With the pulse train determined to consist of 7 operations, we add
two helping pulses, each in the form of Eq.~(\ref{eq32}), to each
operation. The parameters of the first helping pulse are varied so
as to maximize the fidelity in Eq.~(\ref{eq42}) and the process is
repeated for the second helping pulse. Following the optimization
of the helping pulses, we turn to the optimization of the time
spacings between the 7 operations. The fidelity we maximize is now
the fidelity of the entire QFT with decay and decoherence taken
into account as given in Eq.~(\ref{eq62}). The resulting optimized
pulse trains are calculated to have an expected fidelity of 84.9\%
and a duration of 453 ps when helping pulses are not employed, and
an expected fidelity of 88.1\% and a duration of 414 ps when
helping pulses are used. The main source of the infidelity is
unintended dynamics, that is excitation of levels other than the
ones intended. This result is revealed comparing the fidelities in
Eq.~(\ref{eq42}) and Eq.~(\ref{eq62}) for each of the individual
operations. We find the decrease in fidelity from unity due to
decoherence and decay is, on average, 0.8\% per operation, while
the decrease due to unintended dynamics is 2.6\% per operation.
This result could have been expected as the minimum absolute
frequency difference of 36.5 $\mu$eV is always comparable to or
smaller than the spectral width of the pulses.

\section{Conclusions} \label{conclusions}

In this work, we demonstrated a scheme for complete quantum
control of a system of two singly-charged QDs coupled by coherent
electron tunneling.  We derived the energy level diagram for the
system and chose the two-level operations realized by the 5
possible Raman transitions through the trion states as the
universal set of gates. In contrast with the commonly used
universal computation scheme of CNOT gates and single-qubit
gates,\cite{Barenco95} we used only two-qubit gates, each
realizable through a few laser pulses. Our choice is advantageous
as single-qubit gates, which can be described as two simultaneous
two-qubit gates, tend to suffer from unintended dynamics in this
system.\cite{Kim10} We demonstrated our scheme by describing the
realization of the two-qubit QFT in the QDM system with a
reasonably high fidelity of 88.1\%. We incorporated light hole
mixing, decay and decoherence in our analysis.

Our results indicate that pulse shaping can substantially increase
the fidelity of quantum operations in optically active QDs, and
that the QDM is a promising platform for multi-qubit quantum
computation. The main limitations for increasing the overall
fidelity of a given two-qubit operation with this scheme are
unintended dynamics, decay and decoherence. The first can be dealt
with through improved pulse shaping. Due to limited computing
power, we chose our pulse search algorithm to optimize each pulse
separately. A better optimization may be achieved through
simultaneous optimization of all 4 pulses of a given operation.
Ameliorating the effects of decoherence and decay can also be
achieved by improved pulse search methods. In our search we
maximized the fidelity expression in Eq.~(\ref{eq42}) that does
not take decay and decoherence into account, rather than the
expression that takes these effects into account in
Eq.~(\ref{eq62}). Using the latter expression, which takes much
longer to evaluate, will increase the fidelity of the overall
operation.

The experimental realization of universal computation in QDMs will
be a major step towards building a QD-based optically-controlled
quantum computer. Future research may look into methods for robust
and effective entanglement generation between distant QDMs,
possibly with multi-photon light, as was suggested for single
QDs.\cite{Chan13,Cohen13} Quantum computations with more than two
qubits may be realized in a quantum network of entangled QDMs
through unitary operations on each of the QDMs, possibly employing
the theory of cluster states.\cite{Raussendorf03}
\appendix

\section{Decomposition of a Unitary Operation to a Product of Two-Level Proper Unitary Operations}
\label{app-A}

In this appendix we show a procedure to decompose a given unitary
operation $U$ to a product of an overall phase factor
$e^{i\alpha}$ and two-level proper unitary operations. The
procedure is given for the case of a 3-state basis, but can be
easily extended to an $N$-state basis with $N>3$.

The first step in the algorithm is to factor out $|U|=e^{i\alpha}$
and proceed with $U'=e^{-i\alpha/3}U$, which is proper unitary.
Let $U'$ be given by
\begin{equation}
U'=\left ( \begin{array}{ccc} a & d & g \\ b & e & h \\
c & f & g
\end{array} \right ).\label{app-A-eq02}
\end{equation}
If $b=0$, we take $U_1$ as the identity matrix and proceed to the
next step. Otherwise we take $U_1$ as the proper unitary matrix
\begin{equation}
U_1=\frac{1}{\sqrt{|a|^2+|b|^2}}\left ( \begin{array}{ccc} a^* & b^* & 0 \\ -b & a & 0 \\
0 & 0 & 1
\end{array} \right )\label{app-A-eq02}
\end{equation}
and have
\begin{equation}
U_1U'=\left ( \begin{array}{ccc} a' & d' & g' \\ 0 & e' & h' \\ c'
& f' & j'
\end{array} \right ).\label{app-A-eq02}
\end{equation}
Then, if $c'=0$, we take $U_2$ as the identity matrix and proceed
to the next step, while if $c'\neq 0$, we define
\begin{equation}
U_2=\frac{1}{\sqrt{|a'|^2+|c'|^2}}\left ( \begin{array}{ccc}
{a'}^* & 0 & {c'}^* \\ 0 & 1 & 0 \\ -c' & 0 & a'
\end{array} \right ),\label{app-A-eq03}
\end{equation}
which is proper unitary, and have
\begin{equation}
U_2U_1U'=\left ( \begin{array}{ccc} 1 & 0 & 0 \\ 0 & e'' & h'' \\
0 & f'' & j''
\end{array} \right )\equiv U_3.\label{app-A-eq04}
\end{equation}
Hence, the sought decomposition is
\begin{equation}
U=|U|^{1/3}U_1^{-1}U_2^{-1}U_3.\label{app-A-eq05}
\end{equation}

\section{The Minimal Universal Set of Operations in the Quantum Dot Molecule}
\label{app-B}

\begin{figure}[t!]
\begin{center}
\includegraphics[width=5cm]{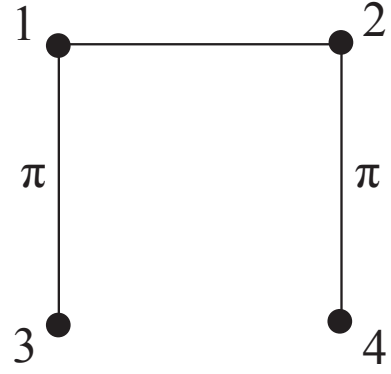}\\
\end{center}
\caption{\label{fig06} Graph representation of the minimal subset
of the possible two-level operations that can still achieve
universal computation. The graph vertices represent levels and its
edges represent possible two-level unitary operations. A
$\pi$-edge represents a $\pm\pi$ rotation operation. The vertices
$1$, $2$, $3$ and $4$ are associated with the states
$|\tilde{S}\rangle$, $|T_0\rangle_x$, $|T_+\rangle_x$ and
$|T_-\rangle_x$, respectively.}
\end{figure}

In this appendix, we prove the minimal universal set of two-level
unitary operations in the quantum dot molecule system is
represented by the graph in Fig.~\ref{fig06}. In the last
paragraph of Sec.~\ref{section2} it was shown that a graph
representing a universal set of operations must be connected and
contain all vertices. Hence, the minimum number of edges is 3. The
edges cannot all be $\pi$ edges since then the number of possible
two-qubit unitary operations attainable with a finite number of
such two-level operations is finite. We conclude the set of
operations represented by the graph in Fig.~\ref{fig06} is
minimal.

This set of operations is universal, since each of the 6 two-level
unitary operations that can appear in the decomposition of an
arbitrary unitary operation $U$ can be written in terms of the 3
operations in the set through use of Eqs.~(\ref{eq25}-\ref{eq26}),
namely
\begin{eqnarray}
R_{12}&=&R_{12},\label{app-B-eq01}\\
R_{13}&=&P_{23}R_{12}P_{23}^\dagger,\label{app-B-eq02}\\
R_{14}&=&P_{21}P_{14}R_{21}P_{14}^\dagger P_{21}^\dagger,\label{app-B-eq03}\\
R_{23}&=&P_{12}P_{23}R_{12} P_{23}^\dagger P_{12}^\dagger,\label{app-B-eq04}\\
R_{24}&=&P_{14}R_{21} P_{14}^\dagger,\label{app-B-eq05}\\
R_{34}&=&P_{14}P_{23}R_{21} P_{23}^\dagger
P_{14}^\dagger,\label{app-B-eq06}
\end{eqnarray}
with $R_{ij}$ and $P_{ij}$ defined in Eq.~(\ref{eq23}) and
Eq.~(\ref{eq24}), respectively.

\section{The Expected Fidelity for Unitary Evolution}
\label{app-C}

In this appendix, we derive the expression for the expected
fidelity of an operation relative to an ideal operation when state
evolution is unitary. The input state $|\psi_{in}\rangle$ is given
by
\begin{equation}
|\psi_{in}\rangle=\sum\limits_{j=1}^4 b_j|j\rangle,\label{eq37}
\end{equation}
while the output state is
\begin{equation}
|\psi_{out}\rangle=\tilde{U}|\psi_{in}\rangle=\sum\limits_{j=1}^8
c_j|j\rangle,\label{eq38}
\end{equation}
with $\tilde{U}$ being the evolution operator. The ideal output
state is given by
\begin{equation}
|\psi_{id}\rangle=U_{id}|\psi_{in}\rangle=\sum\limits_{j=1}^4
d_j|j\rangle,\label{eq40}
\end{equation}
where $U_{id}$ is the ideal evolution operator. The fidelity of
the output state relative to the ideal output state is calculated
as
\begin{equation}
\mathcal{F}=\langle\psi_{id}|\mathrm{Tr}_{T}(|\psi_{out}\rangle\langle\psi_{out}|)|\psi_{id}\rangle,\label{eq39}
\end{equation}
where we trace over the trion states. Plugging Eqs.~(\ref{eq38})
and (\ref{eq40}) in Eq.~(\ref{eq39}), we find
\begin{equation}
\mathcal{F}=\left |\sum\limits_{j=1}^4 c_j^*d_j\right
|^2.\label{eq41}
\end{equation}
Averaging Eq.~(\ref{eq41}) over all possible input states as in
Ref.~\onlinecite{Piermarocchi02}, the expected fidelity of the
operation is obtained in the form of Eq.~(\ref{eq42}).

\section{The Expected Fidelity for Non-Unitary Evolution}
\label{app-D}

In this appendix, we derive the expected fidelity of an operation
relative to an ideal operation when state evolution is
non-unitary. Let the initial state be a pure state
$|\psi_{in}\rangle$. This state evolves to the density matrix
$\rho$ with the evolution written in terms of effect operators
$A_k$ as
\begin{equation}
\rho=\sum\limits_{k=1}^{N_1}A_k|\psi_{in}\rangle\langle\psi_{in}|A_k^\dagger.\label{app-C-eq01}
\end{equation}
The fidelity of $\rho$ in Eq.~(\ref{app-C-eq01}) relative to an
ideally evolved state $|\psi_{id}\rangle=U_{id}|\psi_{in}\rangle$
is given by
\begin{eqnarray}
\mathcal{F}=\sum\limits_{k=1}^{N_1}|\langle\psi_{in}|A_k^\dagger
U_{id}|\psi_{in}\rangle|^2.\label{app-C-eq02}
\end{eqnarray}
When each term in Eq.~(\ref{app-C-eq02}) is averaged over all
possible input states, we find, using
Eqs.~(\ref{eq42}-\ref{eq43}), the expected fidelity is
\begin{eqnarray}
\overline{\mathcal{F}}=\frac{1}{20}\sum\limits_{k=1}^{N_1}\sum\limits_{i=1}^4\sum\limits_{j=1}^4\biggl(I_{ii}^{(k)}I_{jj}^{(k)}+|I_{ij}^{(k)}|^2\biggr),\label{app-C-eq03}
\end{eqnarray}
where $I_{ij}^{(k)}=\langle i|A_k^\dagger U_{id}|j\rangle$.
Changing the order of summation in Eq.~(\ref{app-C-eq03}) and
using Eq.~(\ref{app-C-eq01}) for the time evolution, we obtain
Eq.~(\ref{eq62}).
\begin{acknowledgments}

This research was supported by U.S. Army Research Office MURI
Award No. W911NF0910406, by NSF Grant No. PHY-1104446 and by ARO
(IARPA, W911NF-08-1-0487). The authors thank L. J. Sham for
helpful discussions.

\end{acknowledgments}

\section*{References}
\bibliography{QDM}
\end{document}